\documentclass[a4paper,10pt]{article}
\usepackage[utf8]{inputenc}
\usepackage{graphicx}
\usepackage{amsmath}
\usepackage{float}
\usepackage[margin=1.0in]{geometry}

\title{Penrose Process in a charged axion-dilaton coupled black hole}
\author{Chandrima Ganguly\\ Murray Edwards College, Cambridge CB$3$ $0$DF, University of Cambridge, UK\footnote{cg$528$@cam.ac.uk}\newline \\Soumitra SenGupta\\Department
of Theoretical Physics\\Indian Association for the Cultivation of Science,\\Kolkata $700 032$,India\footnote{tpssg@iacs.res.in}}

\begin{document}

\maketitle

\begin{abstract}
Using the Newman-Janis method to construct the axion-dilaton coupled charged rotating black holes, we show that the energy extraction from such black holes
via Penrose process 
takes place from the axion/Kalb-Ramond field energy responsible for rendering the angular momentum to the black hole. Determining
the explicit form for the Kalb-Ramond field strength, which is argued to be equivalent to spacetime torsion, we demonstrate
that at the end of the energy extraction process, the spacetime becomes torsion free with a spherically symmetric 
non-rotating black hole remnant.
\end{abstract}

\section{Introduction}
The energy extraction process from the black hole following Penrose process gave rise to a new understanding of black hole 
mechanics. Thermodynamic principles associated with black hole geometry gained enormous interest in
recent times in the context of string theory where different kinds of black holes have been investigated in the light of this
principle. Newman and Janis \cite{newmanjanis}, \cite{newmanjanis2} showed that different kinds of inequivalent black holes can be obtained from known solutions, 
for example from Swarzschild black hole , the Kerr solution  can be obtained, while the Kerr Newman metric emerges from the Reissner Nordstrom 
geometry using Newman-Janis prescription.
\paragraph{}
In an alternative approach in string theory, Sen \cite{senhassan},\cite{sen} showed that using T-duality in the low energy string effective action, 
one can generate inequivalent black hole solutions such as charged black hole solutions from uncharged black hole solutions.
After the development of dilaton and axion-dilaton solutions \cite{garfinkle},\cite{yazaddil}, it was shown by \cite{yazad} that
the Newman Janis method can be used to develop the charged axion-dilaton solutions from a pur  dilaton coupled solution, where the axion field plays
the role of the angular momentum parameter of the black hole.
\paragraph{}
It is well known in a string inspired scenario that in four space-time dimensions, the massless axions are dual to the third rank field strength of a second rank antisymmetric tensor field
( known as Kalb-Ramond field )  which appears
in the massless sector of closed bosonic string theories. While this KR field can be interpreted as an external gauge field defined on a Riemannian manifold, there is an alternative viewplint 
where the   third rank anti-symmetric field strength corresponding to the second rank KR field can also
be identified with space-time torsion \cite{hammond,ssg} implying a non-Riemannian geometric structure where the torsion field is completely antisymmetric in its all the
the three indices. Either of these interpretations leads to the same action for the low energy supergravity sector of string theory. 
It is therefore interesting to study the role of the dilaton (whose vacuum expectation value determines the string coupling) as well as the role of the axion in the energy extraction 
process from the black hole via Penrose process.
\paragraph{}
In this work we try to address this question in the context of string inspired axion-dilaton scalar coupled black hole solution obtained
via Newman-Janis prescription. We shall show that, starting from a space-time endowed with non-vanishing KR-energy density ( in other words non-vanishing torsion ), the 
energy extraction process eventually results into a static non-rotating dilaton-coupled black hole where at the end of the energy extraction process, 
axion/KR field strength becomes zero leading to a static Riemannian space-time. 

\section{Axion-dilaton coupled gravity}

The equations of motion for the Einstein-Maxwell axion-dilaton coupled gravity can be obtained from the action,
\begin{equation}
 \mathcal{S} = -\frac{1}{16 \pi} \int d^4x \sqrt{-g} \left(R- 2 \partial _{\mu}\Phi\partial ^{\mu}\Phi - \frac{1}{2} e^{4\Phi} \partial _{\mu}\Theta \partial^{\mu} \Theta + e^{-2\Phi} F_{\mu\nu}F^{\mu \nu} + \Theta F_{\mu \nu} \tilde{F}^{\mu \nu}\right)
\end{equation}
where $R$ is the Ricci scalar with respect to the space time metric $g_{\mu \nu}$ while $F_{\mu \nu}$ and $\tilde{F_{\mu \nu}}$ are the Maxwell tensor and its dual. Here $\Theta$ and $\Phi$ represent 
the axion and the dilaton fields respectively.
Without trying to obtain the solution from the action directly, we first focus on pure dilaton-coupled gravity.

\paragraph{}
The solution to the pure dilaton coupled gravity has been obtained in \cite{garfinkle},\cite{yazaddil}. It resembles the Reissner Nordstrom solution with dilaton scalar and is 
spherically symmetric. This can be written as,
\begin{equation}
 ds^2 = \left(\frac{1-\frac{r_1}{r}}{1+\frac{r_2}{r}}\right)dt^2-\left(\frac{1-\frac{r_1}{r}}{1+\frac{r_2}{r}}\right)^{-1} dr^2-r^2\left(1+\frac{r_2}{r}\right)(d\theta ^2 + \sin ^2\theta d\phi ^2)
\end{equation}
where the dilaton $\Phi$ is given as,
\begin{equation}
 e^{2\Phi}= \frac{1}{1+\frac{r_2}{r}}
\end{equation}
and the radial component of the electromagnetic potential is,
\begin{equation}
A = \frac{-\frac{Q}{r}}{1+\frac{r_2}{r}}
\end{equation}

\paragraph{}
In analogy to the work done by Newman-Janis \cite{newmanjanis},\cite{newmanjanis2} in obtaining the axisymmetric rotating black hole Kerr solution from the spherically symmetric Schwarzschild 
solution, we can now construct the axisymmetric black hole solution from the spherically symmetric solution obtained 
from dilaton-coupled gravity.
This algorithm works by first transforming the metric to the outgoing Eddington-Finkelstein coordinates and then inverting it to express it in the form of its null tetrad vectors as follows,
\begin{equation}
 g^{\mu\nu}= l^{\mu}n^{\nu} + l^{\nu} n^{\mu} - m^{\mu}\overline{m}^{\nu} - m^{\nu} \overline{m}^{\mu}
\end{equation}
The null tetrad thus obtained is then made to undergo the complex transformation which introduces a parameter '$a$'. This transformation is such that the resulting metric is a real 
function of complex arguments $r$ and $u$ which are given as,
\begin{equation}
r \rightarrow r+ ia\cos \theta
\end{equation}
\begin{equation}
 u \rightarrow u - ia\cos \theta
\end{equation}
\begin{equation}
\theta \rightarrow \theta,\;\; \phi \rightarrow \phi
\end{equation}

Some algebraic manipulations yields the metric,
\begin{equation}
ds^2 = -\left(1-\frac{2Mr}{\tilde{\Sigma}}\right)dt^2 -\frac{4aMr}{\tilde{\Sigma}}\sin ^2 \theta dt d\phi -
\small \frac{\tilde{\Sigma}}{\Delta}(dr^2+\Delta d\theta ^2) +\left \{r(r+r_2)+a^2+\frac{2Mra^2 \sin ^2 \theta}{\tilde{\Sigma}}\right\}\sin ^2 \theta d\phi ^2
\end{equation}

where, $\Delta = r(r+r_2) -2Mr + a^2$ and $\tilde{\Sigma}=r(r+r_2) + a^2 \cos ^2 \theta$
Comparing with the solution obtained from \cite{sen}, we have the electromagnetic potential as,
\begin{equation}
A=-\frac{Qr}{\tilde{\Sigma}}(dt-a^2 \sin \theta d\phi)
\end{equation}
The dilaton field $\Phi$ is
\begin{equation}
e^{2\Phi}=e^{2\Phi _0}\left(\frac{r^2 + a^2 \cos ^2 \theta}{r(r+r_2) + a^2 \cos ^2 \theta}\right)
\end{equation}
where the constant multiplicative term $\Phi _0$ is defined as the asymptotic value of the dilaton.\\
The axion field $\Theta$ is given by,
\begin{equation}
\Theta = \frac{Q^2}{M} \frac{a \cos \theta}{r^2 + a^2 \cos ^2 \theta}
\end{equation}
Here $r_2 = \frac{Q^2}{M} e^{2\Phi _0}$ .
Thus we see that with the aid of a co-ordinate transformation we are able to arrive at a solution with non-vanishing axion field. This field is given by the parameter of 
the transformation '$a$' and its effect is to render the black hole with non-zero angular momentum and thereby making it a rotating black hole.

\paragraph{}

The non-vanishing components of the third rank antisymmetric dual Kalb-Ramond (KR) field strength $H_{\mu \nu \alpha}$  can now be determined from the expression of the
axion field $\Theta$ using the duality relation. These turn out to be,

\begin{gather*}
 H_{023}=\frac{\{r(r+r_2)+a^2\cos ^2 \theta -2Mr\}[\{r(r+r_2)+a^2 \cos ^2 \theta\}\{r(r+r_2)+a^2\} +2Mra^2 \sin ^2 \theta]-(2Mra\sin ^2 \theta)^2}{\{r(r+r_2)+a^2\cos ^2 \theta\}^2}
 \\ \nonumber  \times \frac{2Q^2 r a \cos \theta}{M(r^2 + a^2 \cos ^2 \theta)^2}\\ \nonumber
  \times \left[\frac{r(r+r_2) -2Mr + a^2 \cos ^2 \theta-2aMr\sin ^2 \theta}{r(r+r_2) + a^2 \cos ^2 \theta}\right]^{-1}
  \\ \nonumber \times \left[\left(\frac{r(r+r_2) +a^2 \cos^2\theta}{r(r+r_2) -2Mr+a^2}\right)\times\{(r(r+r_2)+a^2 \cos^2 \theta)(r(r+r_2)+a^2)+2Mra^2\sin^2 \theta\}\right]^{-1}
\end{gather*}

and

\begin{gather*}
H_{031}= \frac{\{r(r+r_2) + a^2 \cos ^2 \theta -2 Mr\}[\{r(r+r_2) + a^2 \cos ^2 \theta\}\{r(r+r_2) + a^2\} + 2Mra^2 \sin ^2 \theta]+(2Mra\sin ^2 \theta)^2}{\{r(r+r_2) + a^2 \cos ^2 \theta\}^2}
\\ \nonumber \times \frac{Q^2}{M}\left[ \frac{a \sin \theta}{r^2 + a^2 \cos ^2 \theta} + \frac{a^2 \sin 2\theta \cos \theta}{(r^2 +a^2 \cos ^2 \theta)^2}\right]
\\ \nonumber \times \left[\frac{r(r+r_2) -2Mr + a^2 \cos ^2 \theta-2aMr\sin ^2 \theta}{r(r+r_2) + a^2 \cos ^2 \theta}\right]^{-1}
  \\ \nonumber \times \left[\left(\frac{r(r+r_2) +a^2 \cos^2\theta}{r(r+r_2) -2Mr+a^2}\right)\times\{(r(r+r_2)+a^2 \cos^2 \theta)(r(r+r_2)+a^2)+2Mra^2\sin^2 \theta\}\right]^{-1}
\end{gather*}
These expressions of the KR field strength determines the energy density of the KR field which acts is the source of the rotational energy of the black hole.

Thus we arrive at the existence of non-spherically symmetric solutions for the Kalb-Ramond field strength corresponding to
this class of rotating black hole solutions where the axion and therefore the Kalb-Ramond field acts as the source of  
rotation of the black hole and thus generates an axisymmetric solution from a spherically symmetric dilaton solution. As
discussed earlier, this can also be interpreted as the torsion in the background spacetime and thus from a Riemannian spacetime
we generate an Einstein-Cartan like spacetime.

\section{Horizon structure}
As we have seen in the previous section, the metric obtained from the Einstein-Maxwell-axion-dilaton gravity resembles the Kerr rotating black hole metric with a modification 
caused by the term $r_2$. This is actually the asymptotic value of the dilaton along with a multiplicative constant.
Thus we have a Kerr solution modified by the presence of the dilaton field. We now study the horizon structure of this metric. 
\paragraph{}
We begin with the co-ordinate singularity arising from $\Delta =0$,
\begin{equation}
r(r+r_2) - 2Mr + a^2=0
\end{equation}
The event horizons are then at
\begin{equation}
r_{\pm}= \left(M- \frac{r_2}{2} \right) \pm \sqrt{\left( M- \frac{r_2}{2}\right) ^2 - a^2}
\end{equation}
The metric shows time translation symmtery as well as axial symmetry, due to the independence of $t$ and $\phi$ in the metric componenets. This gives rise to a timelike Killing vectors
$\tau^{\mu}$ and a Killing vector arising out of the axial symmetry $\eta^{\mu}$.
The inner product of the timelike Killing vector gives us,
\begin{equation}
\tau _{\mu} \tau ^{\mu} =1- \frac{2M r}{r(r + r_2) + a^2 \cos ^2 \theta}
\end{equation}
This quantity becomes positive at the event horizon and like the case of the Kerr black hole it becomes zero at a hypersurface at a distance greater than the radius of the event horizon. 
Let this radius be $r_e$ and it is given by,
\begin{equation}
 r_e = \left(M - \frac{r_2}{2}\right) \pm \sqrt{\left(M- \frac{r_2}{2}\right)^2 - a^2 \cos ^2 \theta}
\end{equation}
The region in between $r_e$ and $r_+$ is known as the ergosphere.

\paragraph{}
The maximum angular velocity that a particle can travel with, due to the rotation of the black hole is that of a photon on the event horizon $r_+$, moving along an equatorial orbit.
We define the angular velocity of this photon to be the angular velocity of the black hole. 
In this case, as the photon is moving along the event horizon, we have the condition,
\begin{equation}
\Delta =0
\end{equation}
Therefore we have
\begin{equation}
r(r+r_2) + a^2 = 2Mr
\end{equation}
Considering a light ray that is being emitted in the $\phi$ direction in the $\theta=\pi/2$ equatorial orbit, 
we have the condition for it to be null as,
\begin{equation}
ds^2 = g_{tt}dt^2 +g_{t\phi}(dtd\phi+d\phi dt) + g_{\phi \phi} d\phi ^2=0
\end{equation}
This yields,
\begin{equation}
\frac{d\phi}{dt}=-\frac{g_{t \phi}}{g_{\phi \phi}}\pm \sqrt{\left(\frac{g_{t\phi}}{g_{\phi \phi}}\right)^2 -\frac{g_{tt}}{g_{\phi \phi}}}
 \end{equation}
Using this equation,we find the angular velocity of the photon to be
\begin{equation}
\frac{d\phi}{dt} =0
\end{equation}
and
\begin{equation}
\frac{d\phi}{dt} = \frac{a}{2M r_+} = \Omega _H
\end{equation}
The zero solution indicates that the photon is not moving at all in this frame. 
The non-zero solution shows the angular velocity with which the photon is being dragged around 
in the same direction as the hole's rotation.
The angular velocity of the event horizon itself is defined as the maximum angular velocity of a particle at the horizon.
This quantity is given as,
\begin{equation}
 \omega _H = \frac{a}{2 M r_+}
\end{equation}

\section{Energy extraction and the Penrose process}

A calculation of conservation of energy and angular momentum for a particle in the ergosphere can lead to an energy extraction process
demonstrated by Roger Penrose in the case of a Kerr black hole. Here we consider the procedure followed in \cite{carroll}. Briefly this shows that a particle, in the ergosphere
breaks up into two parts such that one part falls into the event horizon of the black hole and the other escapes out into the 
external universe, in a manner in which, the escaping particle can be shown to have more energy than the original particle
before it breaks apart into two fragments. This excess energy is said to be gained from the rotational energy of the black hole. By repeating this process 
again and again the black hole slows down gradually until the rotation stops altogether and the Kerr black hole becomes a Swarzschild
black hole.

\paragraph{Estimation of energy extraction}

Assume that a particle on entering the ergosphere breaks up into two particles $A$ and $B$. Before breaking up, the four momentum of the whole particle was $p^{(0)\mu}$, and the energy 
was $E=-\tau _{\mu} p^{(0)\mu}$. This energy is positive and conserved along its geodesic. When the particle breaks up into smaller particles, then the four momentum and energy are conserved.
 \begin{align}
  p^{(0)\mu}=p^{(A)\mu}+p^{(B)\mu} \\
  E^{(0)}=E^{(A)}+E^{(B)}
 \end{align}
Here $p^{(A)\mu}$ and $p^{(B)\mu}$ are the four momentums of the two constituent particles and $E^{(A)}$ and $E^{(B)}$ are the corresponding energies.
From the above equations the following analysis can be made.
If the momentum of the second particle be such that its energy is negative, Penrose showed that the initial momentum can be arranged so that afterwards a geodesic trajectory 
can be followed from the Killing horizon back into the external universe. Energy still remains conserved along this path and we have,
\begin{equation}
 E^{(A)}>E^{(0)}
\end{equation}
This implies that the energy with which the first particle leaves the Killing horizon is more than the energy with which it
entered. This energy extraction has come from the rotational energy of the black hole, which in effect originates from the
energy density of the axion/KR field.

\paragraph{}
We now define a new Killing vector, taking into account the modification caused by the dilaton field, as,
\begin{equation}
 \chi ^{\mu} = \tau ^{\mu} + \left(1 - \frac{r_2}{4M}\right)^{-1} \Omega _H \eta ^{\mu}
\end{equation}
For a particle $B$ which crosses the event horizon moving forward in time
\begin{equation}
 p^{(B)\mu} \chi _{\mu} <0
\end{equation}
Using the definitions of $E$ and $L$ as $E=-\tau_{\mu}p^{\mu}$ and $L=\eta_{\mu} p^{\mu}$, we get,
\begin{equation}
 L^{(B)}<\left(1 - \frac{r_2}{4M}\right) \frac{E^{(B)}}{\Omega _H}
\end{equation}
As $E^{(B)}$ is taken to be negative and $\Omega_H$ , the black hole's angular momentum, is positive, the particle must 
have a negative angular momentum. In other words it must be moving against the hole's rotation.
Once the particle $A$ escapes out into the external universe and the particle $B$ falls into the event horizon, 
the energy and angular momentum of the black hole are changed by the negative contributions of the particle $B$ that has 
fallen into it.
Thus we have,
\begin{equation}
 \delta M = E^{(B)}
 \end{equation}
\begin{equation}
 \delta J= L^{(B)}
\end{equation}
The total angular momentum of the black hole $J$ is given by,
\begin{equation}
 J= Ma
\end{equation}
Thus equation $(26)$ becomes 
\begin{equation}
 \delta J<\left(1 - \frac{r_2}{4M}\right)\frac{\delta M}{\Omega _H}
\end{equation}
To find the limit of energy extraction in the case of a Kerr black hole, a quantity known as the irreducible mass is defined
as follows,
\begin{equation}
 M_{irr} ^2 = \frac{A}{16 \pi}
\end{equation}
where $A$ is the area of the event horizon.
$\delta(M_{irr} ^2)$ can shown to be always greater than zero.

\paragraph{}

We find the area of the black hole at the event horizon by defining a constant-time hypersurface at $r=r_+$. 
The metric of this hypersurface is,
\begin{equation}
d\Lambda ^2= \tilde{\Sigma}_+ ^2 d\theta ^2 + \left(\frac{2Mr_+}{\tilde{\Sigma}}\right)^2 \sin ^2 \theta d\phi ^2
\end{equation}
The area is then calculated from the relation
\begin{equation}
dA= \sqrt{g_{\theta \theta} g_{\phi \phi}} d\theta d\phi
\end{equation}
which gives,
\begin{equation}
A= 2Mr_+ \int _0 ^{\pi} \sin \theta d\theta \int _0 ^{2\pi} d\phi= 8\pi M r_+
\end{equation}
When the Kerr parameter '$a$' goes to zero, we get the area of the event horizon of a charged dilaton black hole, 
\begin{equation}
 A_{dilaton}= 16 \pi M \left(M - \frac{r_2}{2}\right)
\end{equation}

The  square of the irreducible mass gives us
\begin{equation}
M_{irr}^2 = \frac{A}{16 \pi} = \frac{M r_+}{2}
\end{equation}
which turns out to be,
\begin{equation}
M_{irr}^2 = \frac{1}{2}\left[\left(M^2 - \frac{Mr_2}{2}\right) + \sqrt{\left(M^2 - \frac{Mr_2}{2}\right)^2 - J^2}\right]
\end{equation}
Thus,
\begin{equation}
\delta(M_{irr} ^2)=\frac{a}{2\sqrt{\left(M-\frac{r_2}{2}\right)^2 -a^2}} \left[\left(1 - \frac{r_2}{4M}\right)\Omega _H ^{-1} \delta M - \delta J\right] 
\end{equation}
The right hand side is greater than zero by the inequality $(33)$ and thus we can say that $M_{irr}$ is the limit to which the
energy of the black hole can be extracted.

\paragraph{}

The maximum amount of energy extraction can be defined by taking into consideration the energy of the black hole that remains after the
energy of the axion from the KR field has been exhausted. In the absence of the axion, we are left with pure dilaton coupled
gravity and the energy that remains is the energy of the charged non-rotating dilaton black hole.\\
From the entropic definition of the irreducible mass (as black hole entropy is proportional to the area of its event
horizon) we can define a quantity similar to the irreducible mass of the charged axion-dilaton black hole. Let the
square of this quantity $M_{dilaton}$ be defined as,
\begin{equation}
 M_{dilaton} ^2 = \frac{A_{dilaton}}{16 \pi}= M \left( M- \frac{r_2}{2}\right)
\end{equation}

The maximum amount of energy extracted before the rotation of the black hole stops can now be found as the 
\begin{equation}
 \Delta M = M_{dilaton}- M_{irr}
\end{equation}
This quantity becomes,
\begin{equation}
 \Delta M = \sqrt{M \left( M- \frac{r_2}{2}\right)} - \sqrt{\frac{M}{2}\left \{ \left(M - \frac{r_2}{2}\right) + \sqrt{\left(M-\frac{r_2}{2}\right)^2 -a^2}\right\}}
\end{equation}

Remembering that  $r_2 = \frac{Q^2}{M} e^{2\Phi _0}$ 
the variation of the energy extraction with the dilaton parameter $r_2$ is plotted,taking $M=2$ and $r_2=1$ as follows. If we take the asymtotic value of the dilaton $\Phi_0$ to be zero then from the above expression we have $Q^2=2$.
\begin{figure}[H]
 \includegraphics[scale=1]{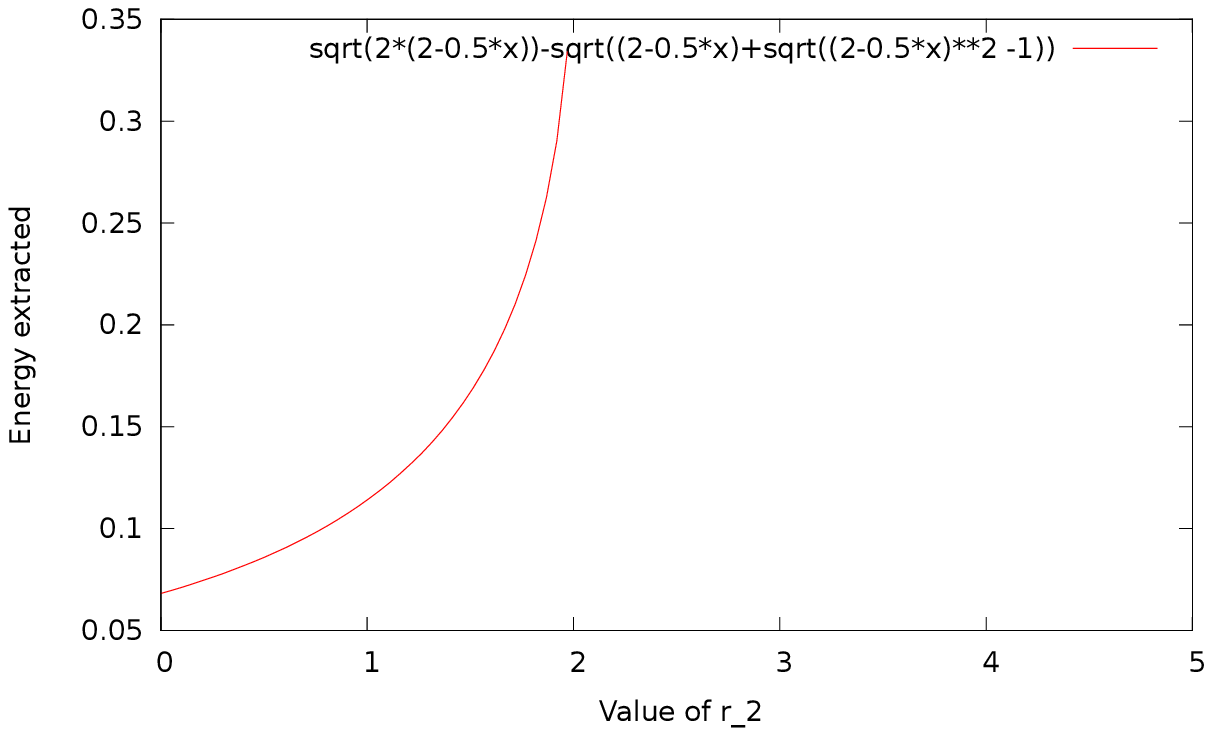}
\end{figure}
As can be seen from the figure the amount of energy extraction reduces with the asymptotic value of the dilaton. The rate of
change of this decrease with the decrease of the dilaton field strength.

\paragraph{}
\begin{figure}[H]
\caption{Variation of energy extraction with axion for $M=2$ and $r_2=1$}
 \includegraphics[scale=1]{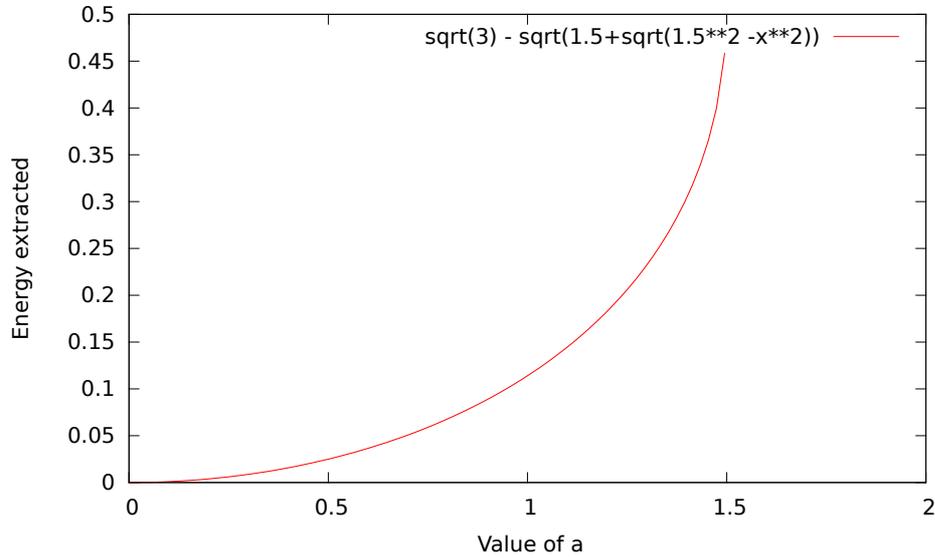}
\end{figure}
Here too, the value of energy extraction reduces with the decrease in the value of the axion as does the rate of energy
extraction with respect to the axion field. When the axion parameter goes to zero, the ergosphere vanishes and the energy
extraction stops, as expected.

\section{Conclusion}

The Newman Janis prescription has been used to generate inequivalent black hole solutions, that is, from the 
spherically symmetric charged dilaton coupled gravity solution to the axisymmetric charged axion-dilaton coupled solution.
\cite{yazad}. The axion in four dimension is dual to Kalb-Ramond field strength which can equivalently be interpreted as space-time torsion and thus  
axisymmetric charged axion-dilaton metric which resembles the Kerr metric, is a black hole solution in a string inspired torsioned space-time.
We have explicitply constructed various components for the torsion from the axion field which in turn is reponsible for rendering angular momentum to the 
black hole. We therefore expect an energy extraction process in a manner similar to the well-known Penrose process. The presence of dilaton also influences the 
space-time geometry and the energy extraction process. We show that at the end of the energy extraction, when the energy from the rotational energy of the black
hole ( or equivalently the energy of the axion or the Kalb Ramond field) has been used up, we are left with a pure charged non-rotating
dilaton black hole. This is, as already mentioned, a spherically symmetric geometry of spacetime and hence we cannot expect
any further energy to be extracted from the dilaton energy by means of the Penrose process.
\paragraph{}
As the black hole entropy is proportional to the area of its event horizon, we can define a quantity known as the irreducible
mass for the axion-dilaton black hole proportional to the area of its outer event horizon. The change in this quantity for the
axion-dilaton black hole can always shown to be positive by defining the inner product of the four momentum of a test
particle with a Killing vector which is actually a linear combination of the timelike and axial Killing vectors in the charged
axion-dilaton black hole spacetime. When the energy of the axion is
fully extracted, we are left with a pure charged dilaton black hole with a remnant entropy proportional to the area of the
event horizon of the dilaton black hole. The difference between these two quantities measures the maximum
amount of energy that can be extracted before the rotation of the black hole finally stops and the energy of the axion is fully extracted. 
The amount of energy extracted is plotted against the parameter measuring the dilaton and the axion field strength. The amount of energy extracted 
as well as the  rate of extraction are found  to decrease with the decrease in the dilaton and the axion field strength. 
\paragraph{}
In this work, we have thus shown that by an energy extraction process similar to the Penrose process, the geometry of the 
spacetime is being altered. The energy is being extracted from the axion of the Kalb-Ramond field (or equivalently the
rotational energy of the black hole). In the presence of the axion, we have an Einstein-Cartan like spacetime,due to the 
equivalent description of the axion as a torsion of spacetime. When the energy of the axion is extracted fully , we are left
with a charged pure dilaton black hole, described via torsionless geometry. This picture is however is a matter of interpretation depending on whether one
would like to interpret the KR field or the equivalent axion as an external tensor field or through the antisymmetric connection of space-time geometry. 

\section{Ackowledgements}
CG would like to thank the Inlaks Shivdasani Foundation and the Cambridge Nehru Bursary for funding her during her studies in the University of Cambridge where part of this work was done.

\end{document}